\documentclass[a4paper]{jpconf}
\usepackage{graphicx}
\usepackage{amssymb,amsmath}
\begin{document}
%
\newcommand{\IJMPB}{{Int. J. Mod. Phys. B} }
\newcommand{\PhC}{{Physica C} }
\newcommand{\PhB}{{Physica B} }
\newcommand{\JS}{{J. Supercond.} }
\newcommand{\IEEEmw}{{IEEE Trans. Microwave Theory Tech.} }
\newcommand{\IEEEas}{{IEEE Trans. Appl. Supercond.} }
\newcommand{\IEEEim}{{IEEE Trans. Instr. Meas.} }
\newcommand{\PRB}{{Phys. Rev. B} }
\newcommand{\IJIMW}{{Int. J. Infrared Millim. Waves} }
\newcommand{\APL}{{Appl. Phys. Lett.} }
\newcommand{\JPCS}{{J. Phys. Chem. Solids} }
\newcommand{\AdP}{{Adv. Phys.} }
\newcommand{\Nat}{{Nature} }
\newcommand{\CM}{{cond-mat/} }
\newcommand{\JpnJAP}{{Jpn. J. Appl. Phys.} }
\newcommand{\PhT}{{Phys. Today} }
\newcommand{\ZETF}{{Zh. Eksperim. i. Teor. Fiz.} }
\newcommand{\JETP}{{Soviet Phys.--JETP} }
\newcommand{\EL}{{Europhys. Lett.} }
\newcommand{\Sci}{{Science} }
\newcommand{\EJPB}{{Eur. J. Phys. B} }
\newcommand{\IJMB}{{Int. J. of Mod. Phys. B} }
%
\title{Effect of nanosize BaZrO$_3$ inclusions on vortex parameters in YBa$_2$Cu$_3$O$_{7-x}$}

\author{E Silva$^1$, N Pompeo$^2$, R Rogai$^2$, A Augieri$^3$, V Galluzzi$^3$, G Celentano$^3$}

\address{$^1$ Dipartimento di Fisica ``E. Amaldi'' and Unit\`a CNISM, Universit\`a Roma Tre, and INFM-CNR ``SUPERMAT",
Via della Vasca Navale 84, 00146 Roma,
Italy}
\address{$^2$ Dipartimento di Fisica ``E. Amaldi'' and Unit\`a CNISM,
Universit\`a Roma Tre, Via della Vasca Navale 84, 00146 Roma,
Italy}
\address{$^3$ ENEA-Frascati, Via Enrico Fermi 45, 00044 Frascati, Roma, Italy}

\ead{silva@fis.uniroma3.it}

\begin{abstract}
We report on the field dependence of the microwave complex resistivity data in YBa$_2$Cu$_3$O$_{7-x}$/BaZrO$_3$ films grown by PLD at various BaZrO$_3$ content. The data, analyzed within a recently developed general framework for the mixed-state microwave response of superconductors, yield the field dependence of the fluxon parameters such as the vortex viscosity and the pinning constant. We find that pinning undergoes a change of regime when the BaZrO$_3$ content in the target increases from 2.5 mol.\% to 5 mol.\%. Simultaneously, the vortex viscosity becomes an increasing function of the applied magnetic field. We propose a scenario in which flux lines are pinned as bundles, and a crossover from dilute point pins to dense $c$-axis correlated defects takes place between 2.5 and 5 mol.\% in the BZO concentration. Our data are inconsistent with vortices occupying mainly the BaZrO$_3$ sites at low fields, and suggest instead that vortices occupy both BaZrO$_3$ sites and interstitials in the YBa$_2$Cu$_3$O$_{7-x}$ matrix, even at low fields.
\end{abstract}

\section{Introduction}
Effective ways to improve pinning properties are essential steps toward practical applications of high-T$_c$ superconductors. The beneficial effect of nanometric inclusions of barium zirconate (BaZrO$_3$, BZO) on the transport properties of YBa$_2$Cu$_3$O$_{7-x}$ (YBCO) received great attention \cite{macmanusNATMAT04, kangSCI06, gutierrezNATMAT07, maiorovNATMAT09}.
The improved dc properties include a greatly enhanced irreversibility line \cite{macmanusNATMAT04, PeurlaPRB07} and pinning force \cite{gutierrezNATMAT07, maiorovNATMAT09}, a reduced field dependence of the critical current density \cite{macmanusNATMAT04,gutierrezNATMAT07, maiorovNATMAT09,augieriIEEE09} and a strong enhancement of the critical current when the field is aligned with the $c$-axis \cite{augieriIEEE09}.

The dc properties probe the long-range vortex motion, that is they probe the motion of the flux lines as a result of their extraction from the pinning sites. As such, dc probes essentially the depth of the pinning wells in the overall pinning landscape.

When a high-frequency, alternate driving current is applied, flux lines undergo very short oscillations (of the order of $\sim$~1nm in the 1-10 GHz range \cite{TomaschPRB88}). Thus, the dynamics that are probed are rather different, and very short-ranged.

We have recently shown that BZO nanoparticles increase the pinning efficacy even at very high driving frequencies, of the order of tens of GHz \cite{pompeoAPL07,pompeoJAP09}: the introduction of a few mol \% of BZO in the target determines a strong decrease of the microwave dissipation. The dissipation monotonously decreases with BZO concentration, from 2.5 mol \% to 7 mol \%., and simultaneously the response becomes increasingly imaginary, indicating an increasingly large reactive component that, in turn, implies a stronger elastic response of flux lines.

In this paper we derive the vortex parameters from the field dependence of the complex resistivity in thin YBCO/BZO films with different BZO concentration, in order to study the changes induced in the vortex pinning and in the vortex structure by the sub-micrometric particles of BZO. We find that a small amount of BZO particles does not significantly change the response, and thus the nature of the vortex pinning and vortex structure. At larger BZO concentrations (from 5 mol.\%) the pinning mechanism changes. Moreover, the data are consistent with two kinds of sufficiently different vortices, that we argue correspond to vortices localized on poorly-conducting BZO sites and vortices localized in the better-conducting YBCO matrix.

\section{Experimental section}
We have measured the microwave response at high frequency ($\sim$48 GHz) with the main purpose of studying the changes in pinning due to the addition of sub-micrometer (as estimated by SEM and optical microscope analysis) BZO particles in the target. 

We used pulsed laser ablation to epitaxially grow YBCO thin films (thickness $d\sim$120$\pm$10 nm) with different amount of BZO content. (001) SrTiO$_3$ (STO) substrates were used. Extensive details of the growth technique and checks of the good sample quality are given elsewhere \cite{galluzziIEEE07,augieriJPCS08}. We mention specifically that the addition of BZO particles in our range, from 2.5 mol.\% to 7 mol.\%, does not deteriorate significantly the sample quality. $T_c$ remained in the range 89.5 K - 90.6 K, and rocking curves from X-ray $\omega$ scans, always with FWHM $\Delta\theta < 0.2^{\circ}$,  showed even improved $c$-axis alignment with increasing BZO content  \cite{pompeoJAP09}. More details and structural analysis can be found in previous publications \cite{galluzziIEEE07,pompeoJPCS08}. Measurements in dc yielded  critical current densities $J_c$ at 77 K consistently with weaker magnetic field dependence as BZO concentration was increased \cite{augieriIEEE09}.

In order to investigate the pinning properties of YBCO/BZO films, a preliminary structural analysis is useful. It has been reported that BZO determines the growth of elongated defects in YBCO/BZO samples \cite{macmanusNATMAT04,kangSCI06,galluzziIEEE07}. This has been directly confirmed in a sample from the same batch of those here studied by transverse TEM measurements \cite{augieriTEM}.
To estimate the density of defects in our YBCO/BZO films we applied a repeated wet-chemical etching method \cite{huijbregtsePRB00}. This method reveals the existence of various kinds of defects, but same shape of the pits indicates same kind of defect. We found that etching produced uniformly distributed pits typical of defects elongated along the $c$-axis. From the same method we evaluated the areal density of such defects. In samples with BZO content (in the target) of 7 mol.\% and 5 mol.\%  we estimated a defect areal density $n_d\sim$~900$\mu$m$^{-2}$ and 600$\mu$m$^{-2}$, respectively, as opposed to the areal density $\sim$10$\mu$m$^{-2}$ in pure YBCO \cite{augieriIEEE09}. Moreover, TEM measurements in films grown, like ours, by pulsed laser deposition showed \cite{peurlaSUST06} that the density of the BZO particles in the films increased with increasing BZO concentration in the target, but the particle size remains practically the same. Thus, we argue that the BZO content in the target produces a related increase in elongated defects in our films.

The field dependence of the complex resistivity was obtained by means of a sapphire dielectric resonator, operating at 47.7 GHz in the TE$_{011}$ mode \cite{pompeoJSUP07}. The superconducting film is a planar wall of the whole resonator, and measurements of the field-induced changes in the quality factor $Q$ and of the resonant frequency $f$ at the fixed temperature $T$ yielded directly the field-induced changes in the complex resistivity, through the standard expression:
\begin{equation}
\frac{\Delta \rho_{1}+\mathrm{i}\Delta \rho_2}{d}= G\left\{\left[\frac{1}{Q(H,T)} -  \frac{1}{Q(0,T)}\right]-2\mathrm{i}\frac{f(H,T)-f(0,T)}{f(0,T)}\right\}
\label{DZeff} 
\end{equation}
In Equation \eqref{DZeff} $G$ is a calculated geometrical factor, $d$ is the film thickness, $B\simeq\mu_0 H$ is the flux density, and we have made use of the thin-film expression \cite{silvaSUST96}. Complications arising from the strong temperature dependence of the substrate permittivity were dealt with according to the method described in \cite{pompeoSUST07}.

Our apparatus allowed to perform measurements for $T>$ 60 K and for fields $\mu_0H<$ 0.8 T. The magnetic field was applied aligned with the $c$-axis. No microwave power dependence was observed for the data here reported. The distribution of microwave currents was such that only the ($a,b$)-plane response was probed. The dc field dependence was found to be reversible, apart from a small field range $\mu_0H \lesssim$~0.05 T.

\begin{figure}[h]
\begin{minipage}{15.5pc}
\includegraphics[width=15.5pc]{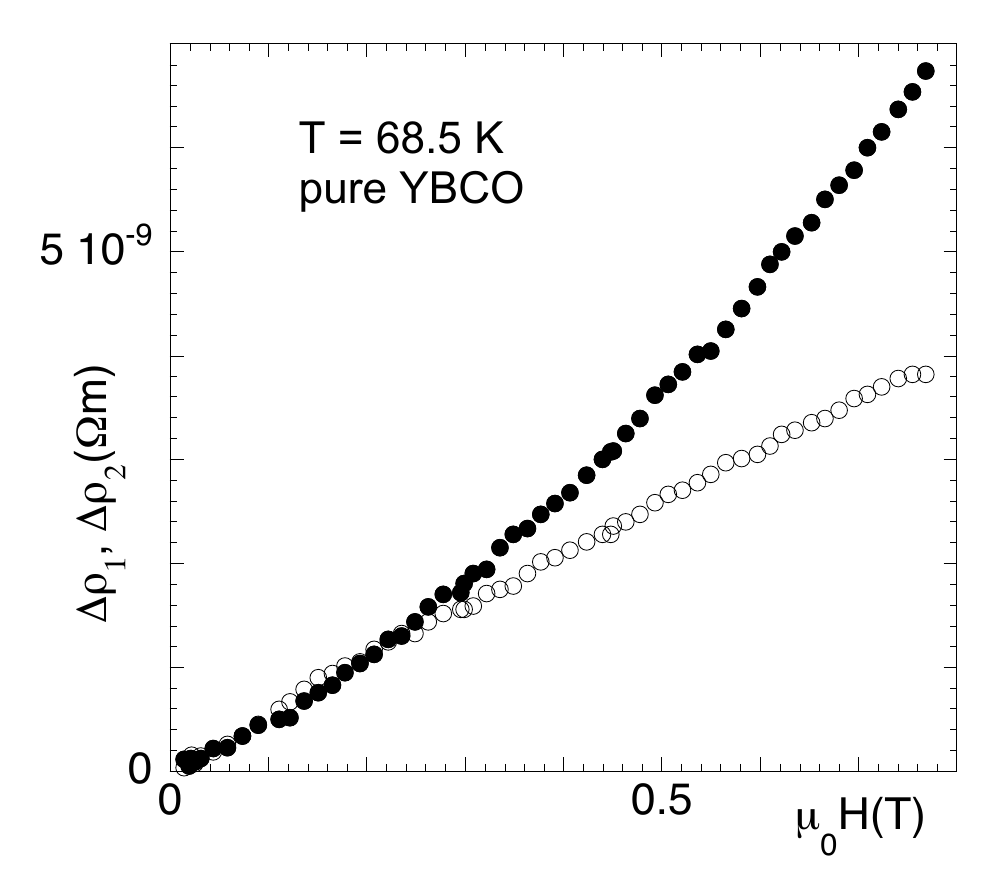}
\caption{\label{figY0}Field-induced complex resistivity shift in a pure YBCO thin film. Full dots: $\Delta\rho_1$. Open circles: $\Delta\rho_2$. To avoid crowding, only 20\% of the data is shown.}
\end{minipage}\hspace{2pc}%
\begin{minipage}{15.5pc}
\includegraphics[width=15.5pc]{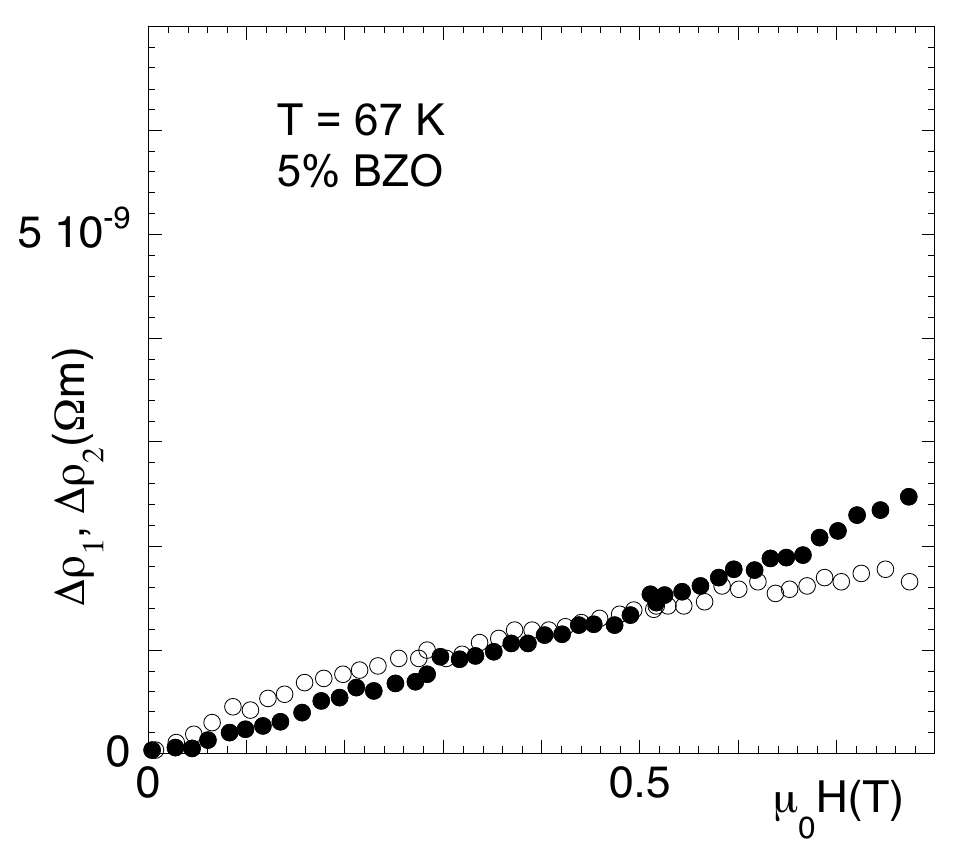}
\caption{\label{figY5}Field-induced complex resistivity shift in a YBCO/BZO at 5 mol.\% thin film. Full dots: $\Delta\rho_1$. Open circles: $\Delta\rho_2$. To avoid crowding, only 20\% of the data is shown.}
\end{minipage} 
\end{figure}

In Figures \ref{figY0} and \ref{figY5} we report typical data for $\Delta\rho_1(H)$ and $\Delta\rho_2(H)$ taken in a pristine YBCO sample and in a YBCO/BZO at 5 mol.\%, respectively. Measurements demonstrate the reversibility of the measurements. It is apparent that the introduction of a small amount of BZO particles changes the response from predominantly dissipative, with $\Delta\rho_1(H)>\Delta\rho_2(H)$, to predominantly reactive, with $\Delta\rho_2(H)>\Delta\rho_1(H)$. Moreover, the field-induced microwave dissipation drops significantly at all fields. Both findings are clear evidence for a very strong pinning, induced by the addition of BZO particles, as it has been discussed previously \cite{pompeoAPL07,pompeoJAP09}.

In order to investigate the changes in pinning mechanisms that give rise to the oberved reduction in dissipation and relative increase in reactance, we use a generalized model previously introduced \cite{pompeoPRB08} to analyize the data and extract the vortex parameter. The model and method are briefly summarized in the following Section. \\

%
%
%

\section{Determination of vortex parameters} 
The field-induced complex resistivity is commonly related to several fundamental vortex parameters, such as the vortex viscosity (sometimes referred to as the vortex drag coefficient) $\eta$ (approximately related to the conductivity $\sigma$ in the vortex cores by $\eta\sim\sigma$), the vortex pinning constant $k_p$, the (de)pinning frequency $\omega_p=2\pi k_p/\eta$. When thermal phenomena are appreciable, creep must be taken into account. In this case another vortex parameter, the ``creep factor" $\chi$ comes into play, and the characteristic frequency $\omega_p$ is replaced by a creep-affected $\omega_0$. There are many models describing the microwave response in terms of some or all of those parameters (see, e.g., \cite{gr,cc,brandt,GolosovskySUST96}). However, it has been shown \cite{pompeoPRB08} that the complex response as predicted by many models can be cast under the following single expression:
\begin{equation}
\label{eqgeneral}
    \Delta\rho(H)=\frac{\Phi_{0}B}{\eta}\frac{\chi+\rmi \frac{\omega}{\omega_0}}{1+\rmi\frac{\omega}{\omega_0}}
\end{equation}
so that significant information can be obtained without assuming a specific model. We have developed a maximum-likelyhood method to obtain the most probable values  $\bar\eta$ for the viscosity, on the basis of the model-independent Equation \eqref{eqgeneral}. To $\bar\eta$ corresponds a value of the creep factor, $\bar\chi$. From $\bar\eta$ and $\bar\chi$ it is straightforward to obtain from Equation \eqref{eqgeneral} the corresponding value $\bar\omega_0$ for the characteristic frequency $\omega_0$. At this stage no specific model has been invoked. One obtains also the most probable value for the pinning constant, $\bar k_p$, within a specific model. Details of the procedure are reported in \cite{pompeoPRB08}.

We discuss our data in terms of the vortex viscosity, $\bar\eta$, and pinning constant, $\bar k_p$. For the latter, we adopted the Brandt model \cite{brandt}, where a phenomenological relaxation time is assumed for vortex escape from a pinning center, instead of an analytical derivation from a specific shape of the pinning potential. However, one must be aware that the model we adopted can be applied only when thermal effects are not too strong, in particular $\chi<\frac{1}{2}$. We confine our attention to the data taken at temperature in the low edge of our range to assure that the constraint is respected, leaving a full temperature analysis to a subsequent work. We focus on the field dependence of $\bar\eta$ and $\bar k_p$ at different BZO content to get insight on the effect of the introduction of BZO particles.

\begin{figure}[h]
\begin{minipage}{16pc}
\includegraphics[width=16pc]{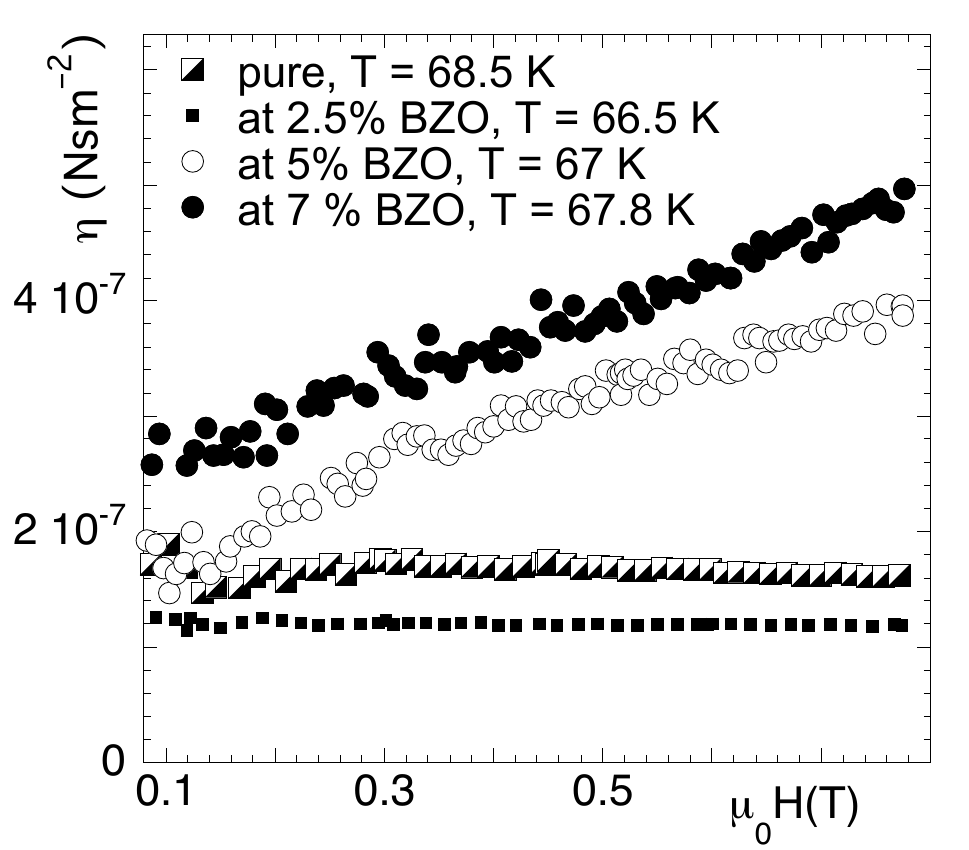}
\caption{\label{figeta}Vortex viscosity $\bar\eta$ as obtained from the data in samples with different BZO content, at similar temperatures. \\ Constant $\bar\eta$ is found only at low BZO content.}
\end{minipage}\hspace{2pc}%
\begin{minipage}{16pc}
\includegraphics[width=16pc]{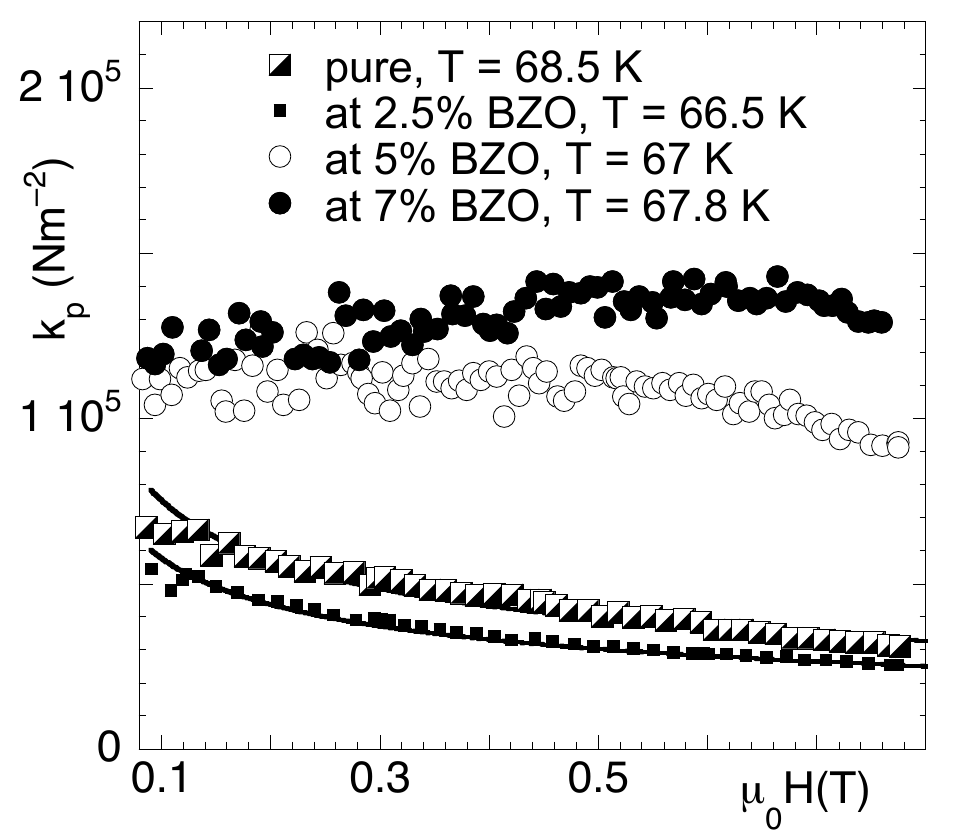}
\caption{\label{figkp}Pinning constant $\bar k_p$ as obtained from the data in samples with different BZO content, at similar temperatures. Continuos lines are fits with $\bar k_p \sim H^{-\alpha}$, with $\alpha=$~0.4.}
\end{minipage} 
\end{figure}

%

\section{Discussion}

Figures \ref{figeta} and \ref{figkp} report the field dependence of the vortex viscosity $\bar\eta$ and the pinning constant $\bar k_p$ as obtained at 66.5 K $-$ 67.8 K in four samples, with different BZO content: 0 mol.\% (pure), 2.5 mol.\%, 5 mol.\% and 7 mol.\%. Scattering is largely increased when both $\Delta\rho_1$ and $\Delta\rho_2$ are close to zero, so that derived data below 0.05 T are somewhat unreliable, and we omit them from the Figures. As it can be seen, both vortex viscosity and pinning constant change their field dependence with increasing BZO content. We discuss first the pinning constant.

First, we note that the low-field absolute values of $\bar k_p$ increase with increasing BZO content, pointing to stronger pinning in YBCO/BZO samples than in pure YBCO. However, this observation should not be confused with evidence for stronger pinning in YBCO/BZO obtained by means of dc measurements: while dc measurements give roughly a measure of the depth of the wells in a pinning potential, the very-short-range dynamics induced by microwave currents probes instead the steepness of such wells. So, from microwave measurements one can draw the conclusion that BZO particle determine the presence of very {\em steep} pinning wells in the overall pinning landscape. In this sense, our observations are complementary to dc observations of increased critical current density $J_c$ in moderate fields on similar samples \cite{augieriIEEE09}.

Interesting information is obtained from the analysis of the field dependence of $\bar k_p$. The pinning constant $\bar k_p(H)$ drops quickly as the field is increased in pure YBCO and in YBCO/BZO at 2.5 mol.\%, while it is approximately constant, with a tendency to increase at low fields, in YBCO/BZO at 5 mol.\% and 7 mol.\%. We argue that the different field dependence marks the crossover between two different pinning regime: at small defect concentration, flux lines are pinned by a dilute array of strong point pins,\footnote{Pins have to be strong in order to be probed at our high measuring frequency: weak point pins do not yield a significant reactive component to nm and sub-nm dynamics.}
 with many vortices per pin. At large defect concentrations, single vortices or small vortex bundles are individually pinned by BZO particles (the analysis of the vortex viscosity, see below, is consistent with the second hypotesis). We note that the array of dilute pins should yield a decrease of the pinning constant as $H^{-\alpha}$, with $\alpha\sim\frac{1}{2}$ \cite{gr,GolosovskySUST96}. The samples at 0 mol.\% and 2.5 mol.\% show a somewhat weaker field dependence, $\bar k_p\sim H^{-0.4}$ (reported as a continuos line in the figure), but not inconsistent with dilute pins model and in agreement with other results obtained at microwave frequencies \cite{powell96}. The sample at 5 mol.\% shows a possible tendency to such a behaviour only above $\mu_0 H=$~0.6T, while the sample at 7 mol.\% does not show any tendency for a decreasing $\bar k_p$.

The vortex viscosity exhibits absolute values well within the range of commonly reported value \cite{GolosovskySUST96}. However, with increasing BZO concentration it exhibits an interesting change of regime, from a nearly constant value in the samples with zero and low BZO concentration, it becomes an {\it increasing} function of the magnetic field at higher BZO concentration. We point out that, while there are several models for a field-decreasing vortex viscosity \cite{Larkin86,Kunchur02}, an increasing viscosity is difficult to understand. Thus, we carefully checked that the field dependence could not be due to artifact of the extraction of the parameters. Following the method described in Reference \cite{pompeoPRB08}, we have found that our data in samples at large BZO concentration are inconsistent with a constant viscosity.

We present a possible scenario for the observed increasing viscosity. According to the analysis of $\bar k_p$, the increasing viscosity might originate from two kinds of flux lines, bound to pinning centers sufficiently different in their electronic structure to give rise to different conductivities in the vortex cores. In this scenario, vortices occupy first the strong pinning centers (BZO-induced correlated defects), where the conductivity is lower and viscosity is consequently smaller. As soon as vortices begin to occupy pure YBCO sites, or sites corresponding to small point pins, the viscosity increases as a consequence of the increased conductivity in the vortex cores. This scenario is quite reasonable due to the very different nature of the elongated defects due to BZO and of the point pins in the YBCO matrix. However, it has a significant consequence: since our data for the viscosity begin to increase immediately upon application of the magnetic field, the proposed scenario requires that flux lines begin to occupy immediately both BZO sites and YBCO matrix. This aspect can be reconciled with the observation of a nearly constant $\bar k_p$ if one assumes that BZO-induced defects extend their pinning effect also to interstitial vortices. As a consequence, the pinning mechanism is necessarily given by pinned bundles of fluxons, possibly of small dimensions, tightly bound together by flux lattice interactions. Each bundle is strongly pinned to a single (or to very few) BZO-induced pinning center. Thus, the beneficial effect of BZO-induced on pinning require a sufficiently rigid flux lattice.

As a final note, we stress that our findings at microwave frequencies are fully consistent with the dc studies on samples grown under the same conditions \cite{augieriIEEE09}: there, it was shown that at 2.5 mol.\% concentration of BZO the effect of pinning did not present any additional directionality along the $c$-axis, suggesting that small BZO concentrations gave rise to some increased point pinning, but did not induced a significant density of $c$-axis correlated defects. By contrast, at and above 5 mol.\% concentration of BZO the critical current density exhibited a giant peak when the field was aligned to the $c$-axis, a further indication of a change of pinning regime.
%

\section{Summary}
Summarizing, we have presented measurements of the microwave complex resistivity in YBCO/BZO thin films at various BZO concentrations, from pure to 7 mol.\%. We have shown that, even on the very short distance probed by the motion of vortices at high frequency, BZO particles strongly increase vortex pinning. With the help of a reliable method to extract the vortex parameters we have shown that a small BZO concentration (2.5 mol.\%) does not change the field dependence of the response with respect to the pure sample. In particular, the data are consistent with a material where a dilute array of point pins exists. By contrast, the analysis of the vortex viscosity and of the pinning constant in samples with high BZO concentration (5 mol.\% and 7 mol.\%) points to a change of the pinning regime, with small flux bundles very strongly pinned by BZO-induced defects. The field-increasing vortex viscosity can be explained by flux lines occupying both $c$-axis correlated defects and interstitial sites in the YBCO matrix.

\end{document}